\documentclass[aps,pre,twocolumn,superscriptaddress,showpacs,amsmath,amssymb]{revtex4-1}
\usepackage[utf8]{inputenc}
\usepackage{amsmath,amsthm}
\usepackage{mathtools}
\usepackage{graphicx}
\usepackage{tikz}
\usetikzlibrary{shapes.arrows}
\usepackage[colorlinks = true,
            linkcolor = blue,
            urlcolor  = blue,
            citecolor = blue,
            anchorcolor = blue]{hyperref}
\usepackage{bbold}
\usepackage{braket}
\usepackage{xcolor}
\usepackage[title]{appendix}
\renewcommand{\L}{\mathbf{L}}

\newcommand{\tr}[1]{\ensuremath{\operatorname{tr}\!{#1}}}

\newcommand{\be}{\begin{equation}}
\newcommand{\ee}{\end{equation}}
\newcommand{\bea}{\begin{eqnarray}}
\newcommand{\eea}{\end{eqnarray}}
\newcommand{\ZZ}{\mathbb{Z}}
\newcommand{\RR}{\mathbb{R}}
\newcommand{\CC}{\mathbb{C}}

\newcommand{\trb}{\mathop{\mathrm{tr}_{0,N+1}}\limits}

\global\long\def\dL{\mathbb{L}}
\global\long\def\dF{\mathbb{F}}

\global\long\def\si{\sigma}

\theoremstyle{thm@}
\newtheorem*{pro}{Proposition}
\newtheorem{lem}{Lemma}
\newtheorem{cor}{Corollary}

\theoremstyle{remark}
\newtheorem*{rem}{Remark}

\begin{document}

\title{Inhomogeneous MPA and exact steady states of boundary driven spin chains at large dissipation}
\author{Vladislav Popkov}
\affiliation{Faculty of Mathematics and Physics, University of Ljubljana, Jadranska 19, SI-1000 Ljubljana, Slovenia}
\affiliation{Bergische Universit\"at Wuppertal, Gauss Str. 20, D-42097 Wuppertal, Germany}
\author{Toma\v z Prosen}
\affiliation{Faculty of Mathematics and Physics, University of Ljubljana, Jadranska 19, SI-1000 Ljubljana, Slovenia}
\author{Lenart Zadnik}
\affiliation{Faculty of Mathematics and Physics, University of Ljubljana, Jadranska 19, SI-1000 Ljubljana, Slovenia}

\begin{abstract}
We find novel site-dependent Lax operators in terms of which we demonstrate exact solvability of a dissipatively driven XYZ spin-$1/2$ chain in the Zeno limit of strong dissipation, with jump operators polarizing the boundary spins in arbitrary directions. We write the corresponding nonequilibrium steady state using an inhomogeneous MPA, where the constituent matrices satisfy a simple set of linear recurrence relations.  Although these matrices can be embedded into an infinite-dimensional auxiliary space, we have verified that they cannot be simultaneously put into a tridiagonal form, not even in the case of axially symmetric (XXZ) bulk interactions and general nonlongitudinal boundary dissipation. We expect our results to have further fundamental applications for the construction of nonlocal integrals of motion for the open XYZ model with arbitrary boundary fields, or the eight-vertex model.
\end{abstract}
\maketitle 

\section{Introduction}

Matrix product ansatz (MPA) is arguably one of the most useful theoretical concepts in statistical and quantum physics of one-dimensional locally interacting systems.
It appears in a diverse variety of contexts, ranging from exact form of the ground state for a certain type of non-integrable spin-1 chains (the so-called AKLT model \cite{AKLT1,AKLT2} of valence bond solids) to exact description of the non-equilibrium steady states of both classical interacting Markov chains (e.g. simple exclussion processes \cite{Derrida,Evans} and driven cellular automata~\cite{Carlos}), as well as Lindblad equation in quantum integrable systems \cite{TP2011,JPAreview}; in special cases it can even describe the full time evolution~\cite{Katja}. Moreover, it enters a general description of the so-called finitely-correlated-states \cite{Werner}, as well as the variational ansatz for a classical simulation of equilibrium and time-dependent quantum states (aka
DMRG-related methods \cite{Vidal,Schollwoeck}). In all known theoretical applications of MPA, the constituent matrices of the ansatz are position independent, and satisfy certain bulk cancellation condition, related to a particular matrix representation of either Yang-Baxter or Zamolodchikov-Faddeev algebra. 

In the context of boundary-driven open quantum systems, i.e. integrable spin chains with Lindblad jump operators that act only on the boundary sites, it is particularly challenging to understand the maximal set of dissipative boundary processes for which the non-equilibrium steady state density matrix can be written exactly. So far, this has only been possible (for  bulk integrable models such as XXZ spin-$1/2$ chain or Fermi-Hubbard model) for the so-called pure-source/pure-sink boundaries, or boundaries which target opposite longitudinal directions~\cite{JPAreview}. Note, though, that a global SU(2) symmetry allows for a more general solvable boundary processes in the isotropic XXX model~\cite{KPS2013}.  It has, however, remained an open question if and how these exact steady state solutions fit into the general framework of integrablity. For example, except in the special case of  dissipatively driven noninteracting models~\cite{fabian}, the solvable dissipatively driven boundaries cannot be generated using the solutions of the ubiquitous reflection equations~\cite{sklyanin}, which constitute the standard framework for generating integrable boundaries in the coherent (nondissipative, Hamiltonian) setting or in the quantum Hamiltonian formulation of classical Markov processes.

In this paper we propose a new direction for a general construction of integrable incoherent boundaries of interacting quantum chains. We construct local  Lax operators that at first sight appear not to be related to the standard solutions of the Yang-Baxter equation for the eight-vertex model~\cite{takhtajan,sklyaninxyz} and have auxiliary dimension which, unlike in the usual analytic scenario, differs from site to site, i.e. auxiliary spaces at different physical sites are manifestly non-isomorphic. These site-dependent Lax operators generate a conserved transfer matrix with an inhomogeneous matrix product structure. 

As a straightforward application of our result we use this mechanism to solve the problem of a boundary driven anisotropic XYZ spin-$1/2$ chain in the limit of strong dissipation (the so-called Zeno regime), where the driving mechanism polarises the boundary-localised degrees of freedom in a fixed direction of arbitrary choice.

The paper is organized in two parts. In the first part we introduce the model and the novel Lax operators and show, how they can be used to construct operators
that commute with the model's Hamiltonian. Our main technical tool is to show the validity of a generalized divergence condition for the Lax operators that guarantees cancellation of unwanted terms in the bulk. The divergence condition appears as an infinite set of recurrence relations that can be solved once the initial seed is provided. Complete analytical ansatz is established rigorously for a special case of the XXZ spin-$1/2$ chain. In the more general case of XYZ model, we only explicitly provide the seed for the recurrence. The complexity of these equations currently only allows us to treat this second case as a numerical scheme. The second part of the letter deals with applications. Here, we introduce the dissipative boundary processes  with arbitrary polarization and, in the limit of strong dissipation, treat them using solutions of the recurrence relations. In the XXZ case we provide the explicit inhomogeneous matrix product form of the non-equilibrium steady state, while in the XYZ case we provide a carefully empirically verified (conjectured) computational recipe for its construction.

\section{Inhomogeneous Matrix Product Ansatz} 

We consider a quantum spin-$1/2$ chain on a $N$-site one-dimensional lattice. Each spin is acted upon by Pauli matrices $\sigma^\alpha\in{\rm End}(\CC^2)$, where $\alpha\in{\cal J}=\{x,y,z\}$. For each $n\in\{1,2,\ldots N\}$ we denote the local one-site operators by $\sigma^\alpha_n=\mathbb{1}_{2^{n-1}}\otimes \sigma^\alpha\otimes\mathbb{1}_{2^{N-n}}$, where $\mathbb{1}_d$ is a $d\times d$ identity matrix. The dynamics that we will consider in this exposition is generated by the anisotropic Heisenberg Hamiltonian
(also known as XYZ model)
\be
H = \sum_{n=1}^{N-1} h_{n,n+1},\qquad h_{n,n+1} = \vec{\sigma}_n \cdot  {\rm J} \vec{\sigma}_{n+1},
\label{eq:Hamiltonian}
\ee
which acts over the total Hilbert space ${\cal H}=(\CC^2)^{\otimes N}$. Here and in the following ${\rm J} ={\rm diag}(J_x,J_y,J_z)$ denotes the diagonal tensor of spin coupling constants (i.e. the anisotropy tensor) and $\vec{\sigma}_n = (\sigma^x_n,\sigma^y_n,\sigma^z_n)$. For ${\rm J}=\mathbb{1}_3$ the model describes the isotropic Heisenberg magnet (XXX model), while ${\rm J}={\rm diag}(1,1,\cos\gamma)$ is  a parametrization of the axially symmetric XXZ model.

We will present a new MPA, by means of which one can construct a positive, semi-definite operator $R=\Omega\Omega^\dagger\in{\rm End}({\cal H})$ that satisfies
\be
[H+\vec{h}_{\rm l}\cdot\vec{\sigma}_1+\vec{h}_{\rm r}\cdot\vec{\sigma}_N,R]=0,
\label{eq:CommutativityRHboundary}
\ee
where $\vec{h}_{\rm l},\vec{h}_{\rm r}\in\RR^3$ are {\em arbitrary} magnetic field polarizations on the left and the right hand side of the spin chain, respectively. The aim is to construct conserved quantities, unrelated to the off-diagonal Bethe ansatz~\cite{ODBA1,ODBA2} that reproduces the integrable hierarchy of the Hamiltonian~\eqref{eq:Hamiltonian} with arbitrary boundary magnetic fields. The main physical application that we shall discuss here is in the context of boundary driven spin chains, also described in the parallel work~\cite{inPrep19}, but one may, in addition, use it for encoding generic (inhomogeneous) conservation laws with a finite spatial correlation structure. 
We start by describing the site-dependent MPA for $\Omega$ and the mechanism responsible for the validity of Eq.~\eqref{eq:CommutativityRHboundary}.

\subsection{Inhomogeneous cancellation mechanism} 

Consider a sequence $\{{\cal A}_n\}_{n=0}^N$ of auxiliary vector spaces with dimensions ${\rm dim}({\cal A}_n)=n+1$ and let $L^\alpha_n,\,I^{}_n \in {\rm Lin}({\cal A}_{n-1}, {\cal A}_{n})$ for $\alpha\in{\cal J}$ be linear maps between them. Denoting $\vec{L}_n=(L_n^x,L_n^y,L_n^z)$ for each $n\in\{1,2,\ldots N\}$, we define Lax operator
\begin{align}
\L_n:=\vec{\sigma}_n\cdot\vec{L}_n=\sum_{\alpha\in{\cal J}}\sigma_n^\alpha L_n^\alpha
\label{eq:Lax}
\end{align}
as an element of ${\rm Lin}({\cal H}\otimes {\cal A}_{n-1},{\cal H}\otimes {\cal A}_{n})$. A sequence of such site-dependent Lax operators will constitute the MPA for the factor $\Omega$ of the operator $R$ satisfying~\eqref{eq:CommutativityRHboundary}. The key to solving this equation is the inhomogeneous Sutherland equation
\bea
[h_{n,n+1},\L_n \L_{n+1}]= i(I_n \L_{n+1}-\L_n I_{n+1}),
\label{eq:sutherland}
\eea
in which the commutator on the left hand side concerns only the operators acting on ${\cal H}$; the equation should be read as
\begin{align}
&\sum_{\alpha,\beta\in{\cal J}}[h_{n,n+1},\sigma_{n}^\alpha\sigma_{n+1}^\beta] L_n^\alpha L_{n+1}^\beta=\notag\\
&\hspace{0.1\linewidth}=i\sum_{\alpha\in{\cal J}}\left(\sigma_{n+1}^\alpha I_n^{} L_{n+1}^\alpha-\sigma_n^\alpha L_{n}^\alpha I_{n+1}^{}\right).
\end{align}
For the ansatz we now set
\be
\Omega=\bra{0}\L_1\cdots \L_N\ket{\psi},
\label{eq:ansatz_r}
\ee
where $\bra{0}\in{\cal A}_0$ and $\ket{\psi}=\sum_{n=0}^N\psi_n\ket{n}$ are boundary vectors that we will identify later. The right boundary vector is inferred from $\bra{\psi}=\sum_{n=0}^N \psi_n^*\bra{n}\in{\cal A}_{N}$ by invoking the duality relation: for $\bra{k}\in{\cal A}_N$, the dual vector is defined through $\braket{k|l}=\delta_{k,l}$. 

Let $\otimes$ denote a  (partial) tensor product of two copies of the auxiliary space that acts as an ordinary matrix multiplication over the physical (quantum) space ${\cal H}$. For $A_n,B_n\in{\rm Lin}({\cal A}_{n-1},{\cal A}_{n})$ and arbitrary $\alpha,\beta\in{\cal J}$ it is defined as
\begin{align}
\big[\sigma_n^\alpha A^{}_n\big]\!\otimes\!\big[\sigma_n^\beta B^{}_n\big]\!:=\!\sigma_n^\alpha\sigma_n^\beta A^{}_n\otimes B^{}_n
\end{align}
and then extended by linearity.
Introducing a two-point Lax operator
\begin{align}
\dL_n=\L^{}_n\otimes\L_n^*:=\sum_{\alpha,\beta\in{\cal J}}\sigma_n^\alpha \sigma_n^\beta\, L_n^\alpha \otimes (L_n^\beta)^*,
\end{align} 
an element of ${\rm Lin}({\cal H}\otimes{\cal A}_{n-1}^{\otimes 2},{\cal H}\otimes{\cal A}_n^{\otimes 2})$, we can now write the MPA for the whole operator $R$:
\begin{align}
R=\bra{0,\bar{0}}\dL_1\ldots \dL_N\ket{\psi,\bar{\psi}}.
\end{align} 
Here  $\ket{\psi,\bar{\psi}}:=\ket{\psi}\otimes(\ket{\psi})^*$, while $(\bullet)^*$ denotes complex conjugation over the auxiliary space and hermitian conjugation over the physical space: $\L_n^* = \sum_{\alpha} (\sigma_n^\alpha)^\dagger (L^\alpha_n)^*= \sum_{\alpha} \sigma_n^\alpha (L^\alpha_n)^*$. As shown in Appendix A, utilisation of the Sutherland equation~\eqref{eq:sutherland} now results in expression
\begin{widetext}
\bea
\begin{gathered}
\big[H+\vec{h}_{\rm l}\cdot\vec{\sigma}_1+\vec{h}_{\rm r}\cdot\vec{\sigma}_N,R\big]=\bra{0,\bar{0}}\dF_1\dL_2  \ldots
\dL_N\ket{\psi,\bar{\psi}}+\bra{0,\bar{0}}\dL_1   \ldots
\dL_{N-1}\dF_N \ket{\psi,\bar{\psi}},\\
\dF_1=[2\,\vec{h}_{\rm l}\cdot\vec{L}_1 + i\,I_1]\otimes \L^*_1 -
\L_1 \otimes [2\,\vec{h}_{\rm l} \cdot\vec{L}^*_1 - i\,I_1],\quad
\dF_N=[2\,\vec{h}_{\rm r}\cdot\vec{L}_N -  i\,I_N]\otimes \L^*_N -
\L_N \otimes [2\,\vec{h}_{\rm r} \cdot\vec{L}^*_N + i\,I_N].
\label{eq:telescopic}
\end{gathered}
\eea
\end{widetext}
We now see that the operator $R=\Omega\Omega^\dagger$, where $\Omega$ is given in \eqref{eq:ansatz_r}, commutes with the Hamiltonian with boundary magnetic fields, i.e. satisfies equation~\eqref{eq:CommutativityRHboundary}, provided that
\be
\bra{0} [2\,\vec{h}_{\rm l}\cdot\vec{L}_1 + i\,I_1] =0,\quad
[2\,\vec{h}_{\rm r}\cdot\vec{L}_N - i\,I_N]\ket{\psi}=0.
\label{eq:boundary}
\ee
In what follows, we will (i) provide unique solutions to the inhomogeneous Sutherland equation~\eqref{eq:sutherland}, thus specifying the MPA, and (ii) show an example of an interesting and nontrivial physical application where the boundary equations~\eqref{eq:boundary} can be solved.

\subsection{Solution of the cancellation mechanism}

A straightforward calculation shows, that Sutherland equation~\eqref{eq:sutherland} is component-wise equivalent to a pair of {\em discrete Landau-Lifshitz} equations
\be
\vec{L}_n \times {\rm J} \vec{L}_{n+1} = \frac{1}{2} \vec{L}_n I_{n+1},\quad
{\rm J}\vec{L}_n \times \vec{L}_{n+1} = \frac{1}{2} I_n \vec{L}_{n+1}.
 \label{eq:recurrence}
\ee
Fixing $\vec{L}_n$ and operators $I_{n},I_{n+1}$, this is an overdetermined set of linear equations for $\vec{L}_{n+1}$. In this paper we will demonstrate (partly prove) and use the following:
\begin{pro}
For a fixed initial datum $\vec{L}_1$ (seed), which depends on two free complex parameters,
there exist a solution to recurrence relations~\eqref{eq:recurrence}, unique up to a choice of basis in each ${\cal A}_n$.
\end{pro}

\subsubsection{Specifying the bases of the auxiliary spaces}

To specify the basis in each auxiliary space, we assume that operators $I_n$ are non-degenerate, choose basis $\{\bra{k;{\cal A}_{n-1}}\}_{k=0}^{n-1}$ of ${\cal A}_{n-1}$, and then define 
\begin{align}
\bra{k;{\cal A}_{n}}:=\bra{k;{\cal A}_{n-1}}I_n, \qquad 0\le k\le n-1
\end{align}
as the first $n$ elements of the basis of ${\cal A}_n$. Next, assuming non-degeneracy of some other Lax component, say $L_n^z$, we define an additional basis vector $\bra{n;{\cal A}_n} := \bra{n-1;{\cal A}_{n-1}}L^z_n$, to get the sequence 
\begin{align}
{\cal A}_0 = \CC\!\bra{0;{\cal A}_0},\quad {\cal A}_n = {\cal A}_{n-1}I_n\oplus \CC\!\bra{n;{\cal A}_n}.
\end{align}
If we interpret the operator $I_n$ as an inclusion map $I_n:{\cal A}_{n-1}\hookrightarrow {\cal A}_n$ we can denote all basis elements by $\bra{k}$, irrespective of the auxiliary space. The entire auxiliary sequence is then embedded in an infinite-dimensional linear space ${\cal A}_\infty={\rm lsp}\{\bra{k}\}_{k=0}^\infty$ and corresponds to
\begin{align}
{\cal A}_0 = \CC\!\bra{0},\quad {\cal A}_n = {\cal A}_{n-1}\oplus \CC\!\bra{n}.
\end{align}
Introducing the dual basis $\{\ket{k}\}$ through the orthogonality relation $\braket{k|l}=\delta_{k,l}$, the operators $L^\alpha_n$ and $I^{}_n$ can now be represented as rectangular $n \times (n+1)$ matrices 
\begin{align}
L_n^\alpha=\sum_{k=0}^{n-1}\sum_{l=0}^{n}L_{n;k,l}^\alpha \ket{k}\!\bra{l},\quad I_n = \sum_{k=0}^{n-1}\ket{k}\!\bra{k},
\label{eq:Lax_components}
\end{align}
where $L^z_{n;n-1,l}=\delta_{n,l}$. Having specified the basis, the boundary vectors of the MPA~\eqref{eq:ansatz_r} are now identified. On the left hand side we simply take the state $\bra{0}$ that, by subsequent action of $L_1^z,L_2^z,\ldots$, generates the entire basis in the embedded space ${\cal A}_\infty$, while on the right hand side $\ket{\psi}=\sum_{n=0}^N\psi_n\ket{n}$, where the complex parameters $\psi_n$ will be fixed by the second 
of the boundary equations~\eqref{eq:boundary}.

\subsubsection{Solutions of the recurrence and connection to integrability}

Solving the nonlinear coupled equation~\eqref{eq:recurrence} for $\vec{L}_1$, at $n=1$ and for arbitrary spin coupling constants $J_\alpha$ we obtain -- up to either trivial or equivalent solutions -- the following two-parametric solution for the seed
($\xi,\eta\in\CC$):
\bea
L^x_1 &=& \begin{pmatrix}
\xi & \frac{\eta}{(\xi^2+\eta^2)(\omega_{xy}\eta^2-1)}\sqrt{r}
\end{pmatrix},\label{eq:seed}\\
L^y_1 &=&  \begin{pmatrix}
\eta & \frac{\xi}{(\xi^2+\eta^2)(\omega_{xy}\xi^2+1)}\sqrt{r}
\end{pmatrix},\label{eq:seed1}\\ 
L^z_1 &=&  \begin{pmatrix}
0 & 1
\end{pmatrix},\label{eq:seed2}
\eea
where $\omega_{\alpha\beta}:=4\,(J_\alpha^2-J_\beta^2)$ and we have denoted
\begin{align}
r = (\xi^2\!+\!\eta^2)(\omega_{xy}\eta^2\!-\!1)(\omega_{xy}\xi^2\!+\!1)(\omega_{xz}\xi^2\!+\!\omega_{yz}\eta^2\!+\!1).
\end{align} 
Using a symbolic computer algebra we have checked, that the overdetermined linear equations~\eqref{eq:recurrence} now generate unique $\vec{L}_n$ for $n=2,3\ldots N$. Each matrix element of any auxiliary Lax component $L_n^\alpha$ 
is of the form $p(\xi,\eta) +  q(\xi,\eta) \sqrt{r}$, where $p,q$ are some rational functions. Unfortunately the complexity of the solution quickly increases with $n$ and we were unable to determine its explicit analytic structure. Hence, for $n>5$ and arbitrary $J_\alpha$ one can only efficiently solve the recurrence equations~\eqref{eq:recurrence} numerically.

Nevertheless, for the special case of XXZ model, where $J_x=J_y=1, J_z=\cos\gamma$, $\gamma\in\RR$ (or $i\RR$), the recurrence (\ref{eq:recurrence}) can in fact be explicitly analytically solved (see Appendix B). After changing the basis by writting
\begin{align}
L^x_n =\frac{1}{2}(L^+_n + L^-_n),\quad L^y_n = \frac{1}{2i}(L^-_n - L^+_n),
\end{align} 
the solution reads
\be
\begin{gathered}
L^z_n = \sum_{k=0}^{n-1} \ket{k}\!\bra{k+1},\\
L^\pm_n= \pm \eta^{\mp 1}\sum_{k=0}^{n-1}\sum_{l=0}^{n}\left(\frac{\pm i}{2 \cos\gamma}\right)^{k-l+1}\!\!\!M_{n;k,l}\ket{k}\!\bra{l},
\label{eq:seedXXZ1}
\end{gathered}
\ee
where we have introduced the following symbols:
\be
\begin{gathered}
M_{n;k,l} = \frac{(\xi-\xi^{-1})\, P_{n,k+1,l}(\cos\gamma)}{(\xi+\xi^{-1})\,\sin\gamma}-\frac{2P_{n,k+1,l-1}(\cos\gamma)}{(\xi+\xi^{-1})\,\cos\gamma},\\
P_{n,k,l}(x)=\sum_{m=0}^l (-1)^m {n-k\choose m}{n-m-1\choose l-m}\,x^{n-2m}.
\label{eq:seedXXZ2}
\end{gathered}
\ee
Here, the free variables $\xi,\eta\in\CC$ provide a different parametrization than those in Eqs.~(\ref{eq:seed}), where the spin coupling constants $J_\alpha$ are arbitrary. For general $\xi,\eta$ we have checked that this solution of the Sutherland equation~\eqref{eq:sutherland} [or, equivalently, the recurrence relations~\eqref{eq:recurrence}] cannot be reduced to any known solution of the Yang-Baxter equation by means of local twists in the auxiliary spaces ${\cal A}_n$. Instead, recombining the Lax component $L^z_n$ and the inclusion operator $I_n$ as $K_n^\pm=\tfrac{i}{2}(I_n\pm2\sin\gamma L_n^z)$ and denoting $q=e^{i\gamma}$, we notice that Eq.~\eqref{eq:recurrence} is equivalent to the inhomogeneous quantum group relations
\begin{align}
\begin{aligned}
K_n^+L_{n+1}^\pm&=q^{\pm 1}L_n^\pm K_{n+1}^+,\\
K_n^-L_{n+1}^\pm&=q^{\mp 1}L_n^\pm K_{n+1}^-,\\
L_n^+L_{n+1}^--L_n^-L_{n+1}^+&=\frac{K_{n}^+K_{n+1}^+-K_n^-K_{n+1}^-}{q-q^{-1}},\\
K_{n}^+K_{n+1}^-&=K_{n}^-K_{n+1}^+.
\label{eq:su2q}
\end{aligned}
\end{align}
These relations define an inhomogeneous analogue of the $q$-deformed spin algebra ${\cal U}_q(sl_2)$ and thus provide a yet-unexplored manifestation of the XXZ spin-$1/2$ chain integrability structure, albeit now with an extensive (in system size) number of generators. Note that, when site-independent, relations~\eqref{eq:su2q} describe the $sl_2$ spin algebra as $q\to 1$ ($XXX$ model). On the other hand, while the matrices $L^\alpha_n$ of our ansatz~\eqref{eq:seedXXZ1} do become elementwise site-independent in this limit, they do not reduce to the standard spin ladder operators. In the following we illustrate the facility of the recurrence scheme~\eqref{eq:recurrence} for solving a boundary driven Lindblad equation in the regime of strong dissipation.

\section{Application: Quantum Zeno limit of the boundary driven XYZ chain} 

We wish to use the ansatz, described in the preceding section, to construct the nonequilibrium steady state (NESS) of the Lindblad equation
\be
\frac{{\rm d}}{{\rm d}t} \rho(t) = -i [H',\rho(t)] +\Gamma\,{\cal D}_{\rm l}[\rho(t)]+\Gamma\,{\cal D}_{\rm r}[\rho(t)],
\label{eq:lindblad}
\ee
at large dissipation strength $\Gamma$, where ${\cal D}_{\mu}[\rho]$, $\mu\in\{{\rm l},{\rm r}\}$,  denote the dissipators at the left and right ends of the chain of $N+2$ sites, which we label by $0$ and $N+1$, respectively. They are of the form
\begin{align}
{\cal D}_{\mu}[\rho]=2k^{}_{\mu}\rho k^\dagger_{\mu}-\{k^\dagger_{\mu}k^{}_{\mu},\rho\},
\end{align}
with the two jump operators 
\begin{align}
k^{}_{\rm l,r}=(\vec{n}'_{\rm l,r} + i \vec{n}''_{\rm l,r})\cdot \vec{\sigma}_{0,N+1}
\end{align}
targeting polarizations $\vec{n}_{\mu}=\vec{n}(\theta_{\mu},\phi_{\mu})$, where 
\begin{align}
\vec{n}(\theta,\phi)=(\sin \theta \cos \phi,\sin \theta\sin \phi,\cos \theta).
\label{eq:unitvectors}
\end{align}
Real vectors $\vec{n}'_{\mu} = \vec{n}(\frac{\pi}{2}-\theta_{\mu},\pi+\phi_{\mu})$ and $\vec{n}''_{\mu} = \vec{n}(\frac{\pi}{2},\phi_{\mu}-\frac{\pi}{2})$, together with $\vec{n}_{\mu}$, form an orthonormal basis of $\mathbb R^3$.
The targeted states of the dissipators are single-site pure states $\rho_{\mu}=|\psi_\mu\rangle \langle\psi_\mu|$, such that 
\begin{align}
{\cal D}_{\mu}[\rho_{\mu}]=0,\qquad {\rm tr}[\rho_\mu\vec{\sigma}]=\vec{n}_\mu,
\end{align} 
i.e. $k_\mu\ket{\psi_\mu}=0$ ($\ket{\psi_\mu}$ should not be confused with the right boundary vector $\ket{\psi}$ of the MPA).
The Hamiltonian is now provided by Eq.~\eqref{eq:Hamiltonian} extended by two sites:
\begin{align}
H'=H+h_{0,1}+h_{N,N+1}.
\end{align}

The problem of constructing the NESS in the limit of strong dissipation (Zeno limit) has been rigorously examined, but not solved in \cite{2018PopkovZenoDynamics}. In the limit $\Gamma\to\infty$, when the unitary part of the dynamical equation~\eqref{eq:lindblad} can be neglected, the NESS should obviously be of the form 
\begin{align}
\rho^{(0)}=\rho_{\rm l}\otimes R\otimes\rho_{\rm r},
\end{align}
where $R$ is some operator acting on the Hilbert space ${\cal H}$ of the internal degrees of freedom, labeled with $1,2,\ldots N$. For large but finite $\Gamma$, we can proceed perturbatively by expanding $\rho_{\infty}=\sum_{k\ge0}\Gamma^{-k}\rho^{(k)}$. Plugging the expansion into the Lindblad equation~\eqref{eq:lindblad}, demanding stationarity ${\rm d}\rho_{\infty}/{{\rm d}t}=0$ and comparing the orders of $\Gamma^{-1}$, we get 
\begin{align}
{\cal D}_{\rm l}[\rho^{(0)}]+{\cal D}_{\rm r}[\rho^{(0)}]=0,
\end{align} 
which is automatically satisfied, as well as a sequence of equations
\be
{\cal D}_{\rm l}[\rho^{(k+1)}]+{\cal D}_{\rm r}[\rho^{(k+1)}]=i [H',\rho^{(k)}],\qquad k\ge 0.
\label{eq:sequence_of_eqs}
\ee
Equations~\eqref{eq:sequence_of_eqs} in particular state that $[H',\rho^{(k)}]$ belongs to the image space of the dissipator (superoperator) ${\cal D}_{\rm l}+{\cal D}_{\rm r}$, which only acts on the boundary degrees of freedom, labeled by $0$ and $N+1$. Since normalization of the density matrix is conserved by the Lindblad equation~\eqref{eq:lindblad} and separately by its unitary part, respectively ${\rm d}(\tr[\rho(t)])/{{\rm d}t}=0$, and $\tr[H',\rho(t)]=0$, this yields additional condition $\trb[H',\rho^{(k)}]=0$~\cite{NJPdecoherence2015}. For $k=0$ it explicitly reads
\be
[H_{\cal D},R]=0,
\label{eq:commutation}
\ee
where $H_{\cal D}$ is the dissipation-projected Hamiltonian that acts on sites $1,2,\ldots N$ and takes the following form:
\be
H_{\cal D}= H+ ({\rm J}\vec{n}_{\rm l})\cdot \vec{\si}_1 +  ({\rm J}\vec{n}_{\rm r})\cdot \vec{\si}_{N}.
\label{eq:dis_proj_ham}
\ee
Although this condition seems rather insignificant at first, it remarkably constitutes the core of our solution to the problem of the strongly boundary driven spin chain.

Indeed, we have arrived at the problem defined in the first section, which can be solved by our ansatz~\eqref{eq:ansatz_r}, if equations \eqref{eq:boundary} are satisfied for the boundary field orientations $\vec{h}_{\rm l}={\rm J}\vec{n}_{\rm l}$ and $\vec{h}_{\rm r}={\rm J}\vec{n}_{\rm r}$. The left boundary equation $\bra{0} [2\,({\rm J}\vec{n}_{\rm l})\cdot\vec{L}_1 + i\,I_1] =0$, in reality a set of two equations for two variables, completely fixes the parameters $\eta$ and $\xi$ in the components $L^\alpha_1$. In the XXZ case, the solution reads
\bea
\eta=-e^{i\phi_{\rm l}}\tan\left(\frac{\theta_{\rm l}}{2}\right),\quad
\xi=\frac{\cos\gamma}{\sin\gamma-1}.
\eea
In the general XYZ model, the solution to the left boundary equation~\eqref{eq:boundary} exists as well and is unique for our choice of bases in ${\cal A}_{0}$ and ${\cal A}_{1}$. Alternatively, we can choose a gauge, different than in Eqs.~\eqref{eq:seed},~\eqref{eq:seed1} and~\eqref{eq:seed2}, in which the seed that generates the solution to the recurrence~\eqref{eq:recurrence} becomes explicitly dependent on the left-edge polarisation axes $\vec{n}_{\rm l}$,
\be
L_1^\alpha=
\frac{1}{2J_\alpha}
\begin{pmatrix}
-i n_{\rm l}^\alpha
& n_{\rm l}^{\prime\,\alpha}- i  n_{\rm l}^{\prime\prime\,\alpha}
\end{pmatrix},
\ee
and satisfies the left boundary equation by construction.

Having specified the parameters, thus fixing the ansatz in the bulk of the system we now turn to the right boundary equation in~\eqref{eq:boundary},
i.e. $[2\,({\rm J}\vec{n}_{\rm r})\cdot\vec{L}_N - i\,I_N]\ket{\psi}=0$, which determines $\ket{\psi}$. Writing $\ket{\psi}=\sum_{n=0}^{N}\psi_n\ket{n}$, with $\psi_0=1$, this is a set of $N$ linear equations for $N$ unknowns $\psi_n$. One solution always exists and seems to be unique for generic values of the boundary angles $\theta_{\mu}$ and $\phi_{\mu}$. In particular cases, for example, for XXZ chain with $\theta_{\rm r}=\phi_{\rm r}=0$, it can easily be computed analytically: 
\be
\psi_n=[i/(2\cos\gamma)]^n.
\ee 
In general, we compute it numerically.

\begin{figure}[ht!]
\centering
\includegraphics[trim={1 0 0 0},clip,width=1\linewidth]{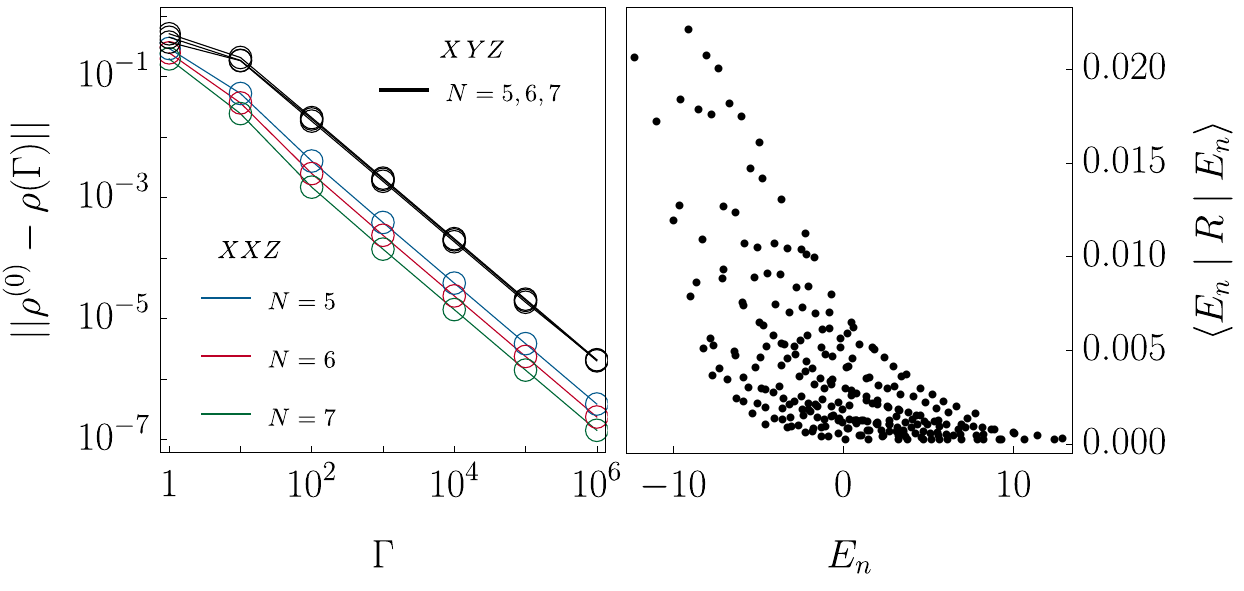}
\caption{Panel (a) shows the difference $\|\rho^{(0)}-\rho(\Gamma)\|$, between our ansatz for NESS in the Zeno limit and the solution to $i[H',\rho(\Gamma)]=\Gamma\,{\cal D}_{\rm l}[\rho(\Gamma)]+\Gamma\,{\cal D}_{\rm r}[\rho(\Gamma)]$, respectively. $\|\bullet\|$ represents the operator norm and $N$ the number of internal sites, i.e. sites not acted upon by the dissipation. Panel (b) shows the scatter plot of eigenvalues of $R$ versus eigenvalues of the dissipation-projected Hamiltonian $H_{\cal D}$ in a generic point where the spectrum of $H_{\cal D}$ is nondegenerate, for $N=8$, indicating functional independence of operators $R$ and $H_{\cal D}$.}
\label{fig:check}
\end{figure}

When unique, the resulting operator $\rho^{(0)}=\rho_{\rm l}\otimes R\otimes\rho_{\rm r}$ indeed reproduces the NESS of the Lindblad equation \eqref{eq:lindblad} in the Zeno limit: (i)~In special cases, where the latter is known analytically~\cite{2012XYtwist}, we find it in complete agreement with our ansatz. (ii)~In generic cases, we resort to comparison with numerically exact NESS, computed via a method proposed in~\cite{2018PopkovZenoDynamics}, which yields equivalence up to the preset numerical precision. (iii)~For finite values of the dissipation strength $\Gamma$, the ansatz $\rho^{(0)}$ converges towards the NESS of the finite-$\Gamma$ Lindblad equation~\eqref{eq:lindblad}, i.e. towards the solution of 
\begin{align}
i[H',\rho(\Gamma)]=\Gamma\,{\cal D}_{\rm l}[\rho(\Gamma)]+\Gamma\,{\cal D}_{\rm r}[\rho(\Gamma)],
\end{align}
as shown in Fig.~\ref{fig:check}, again indicating that the ansatz is correct. The right-hand-side plot on Fig.~\ref{fig:check} also shows that operators $R$ and $H_{\cal D}$ are functionally independent, i.e. $R\neq f(H_{\rm D})$ for at least a piece-wise smooth function $f$, in turn implying nontriviality of our ansatz.

Note that there are also cases, in which the right boundary vector $\ket{\psi}$ of the MPA is not unique. We hypothesize this to happen in measure-zero subset of the parameter space.
Even in this case, however, we find that the Zeno NESS is correctly reproduced by our ansatz for a specific choice of the right boundary vector. Resolving this issue analytically 
requires considering higher orders $\rho^{(k)}$ of the perturbative expansion, which is out of our present scope.

\begin{figure}[ht!]
\centering
\includegraphics[trim={0 0 0 0},clip,width=1\linewidth]{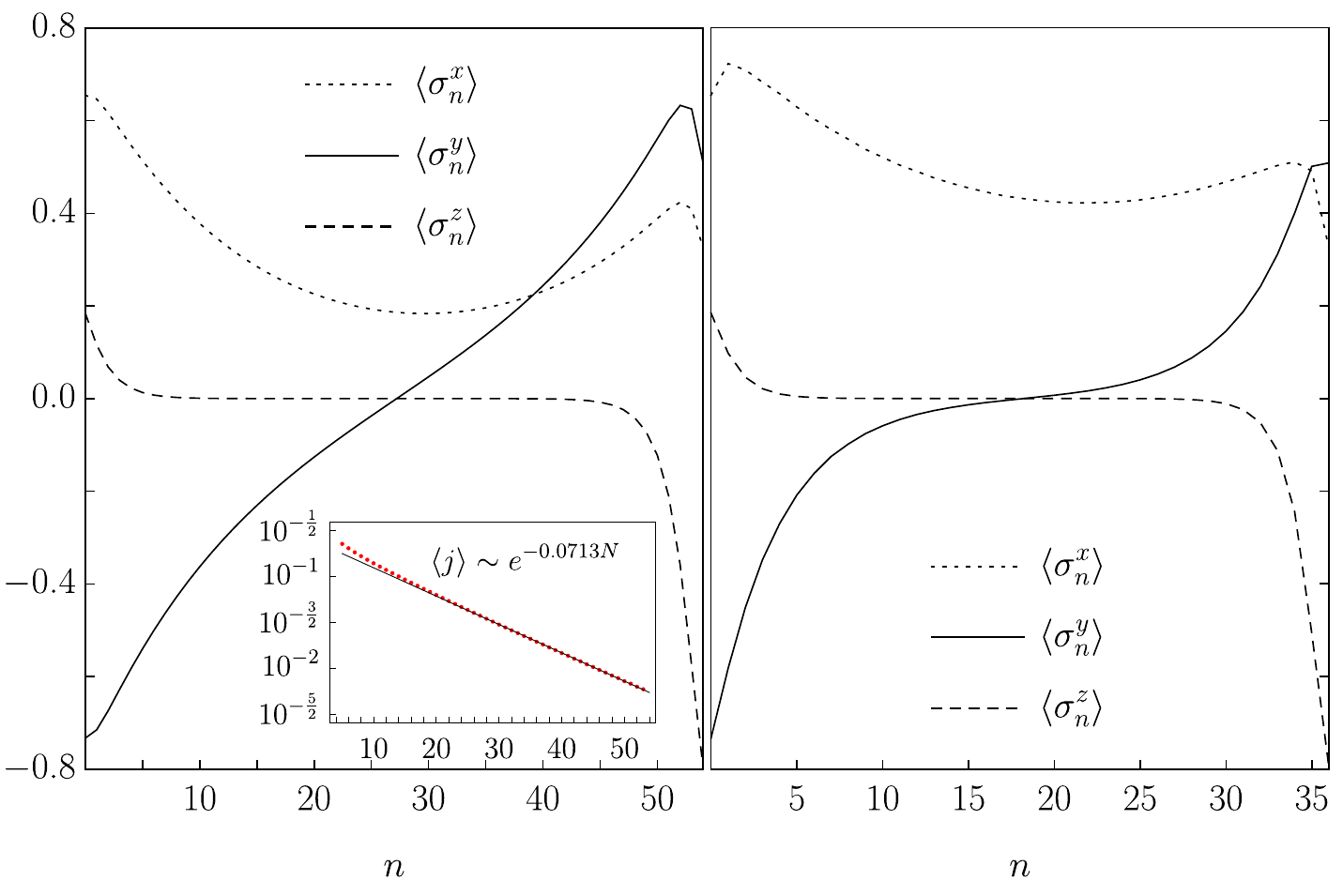}
\caption{Profiles of magnetization in XXZ spin chain (a) and XYZ spin chain (b). The inset on the panel (a) shows exponential decay of the current with system size in the XXZ case. This is a generic example of our problem, parameters being $\phi_{\rm l}=\sqrt{3}\pi$, $\theta_{\rm l}=(1-\sqrt{5}/4)\pi$, $\phi_{\rm r}=\sqrt{5}\pi/7$ and $\theta_{\rm r}=(7-\sqrt{5})\pi/6$. In the XXZ case $\gamma=(\sqrt{5}-1)\pi/8$ and in XYZ case $J_x=13/10$, $J_y=6/5$, $J_z=1$. System sizes (without the sites on which the jump operators act) are $N=53$ and $N=35$, respectively.}
\label{fig:profiles_scaling}
\end{figure}

\begin{figure}[ht!]
\centering
\includegraphics[trim={10 0 0 0},clip,width=1\linewidth]{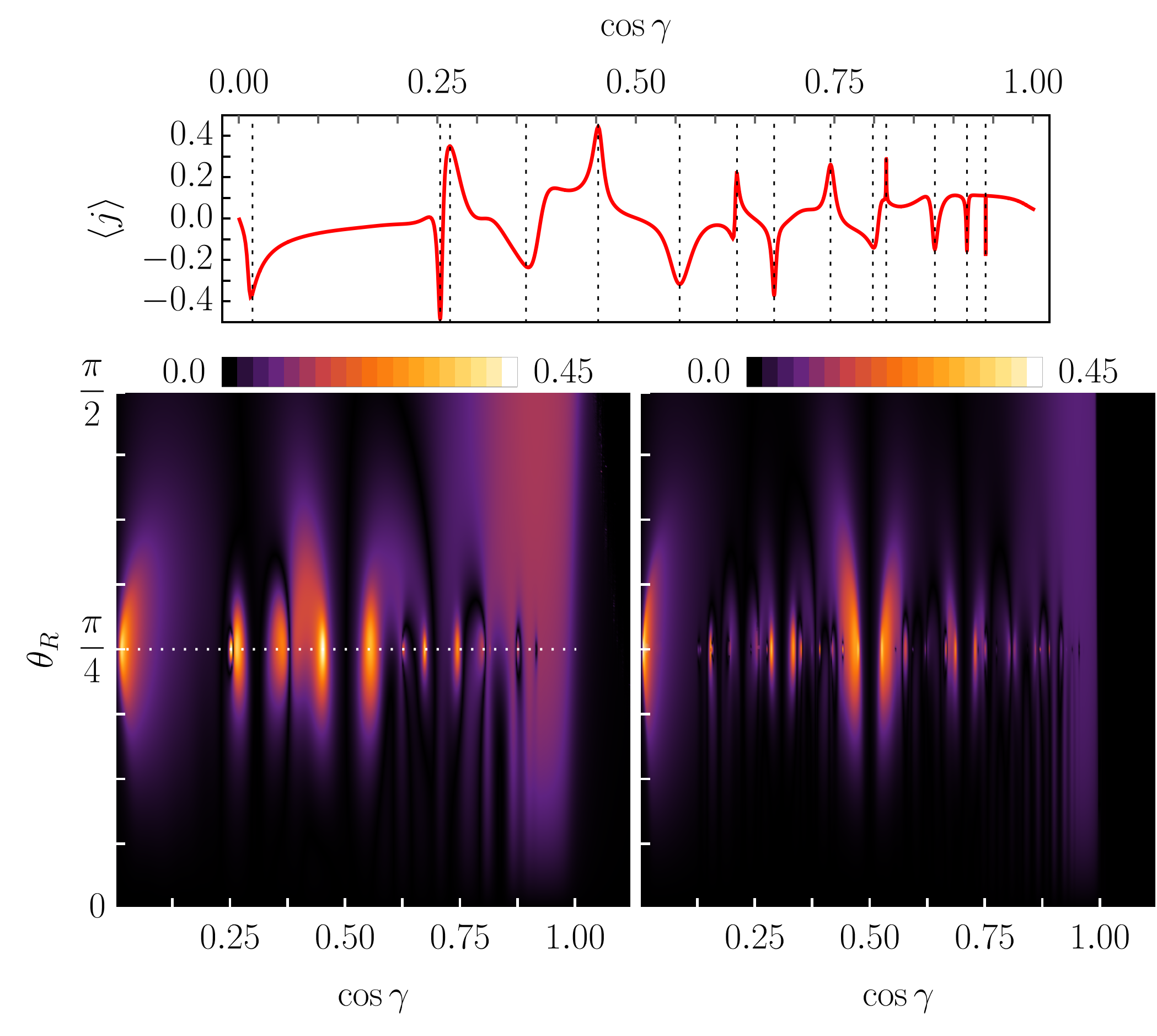}
\caption{The lower two plots show absolute value of the spin current average $\langle j\rangle$ as a function of $\theta_{\rm r}$ and the anisotropy $\cos\gamma$ in the Zeno regime of the XXZ chain for $N=12$ (b) and $N=24$ (c). The upper diagram (a) shows the cross section at $\theta_{\rm R}=\pi/4$ in the gapless regime for $N=12$ (dotted white line on the bottom left plot). Other parameters are $\phi_{\rm l}=\pi/4$, $\theta_{\rm l}=\pi/4$, $\phi_{\rm r}=(3/4+1/17)\pi$. The resonance peaks in the current average (top plot) are located at some of the points characterized by $\cos\left([2\pi m+(\phi_{\rm r}-\phi_{\rm l})]/[m_0+1]\right)$,  where $m_0=N,N-2,N-4,\ldots$ and $m=0,1,\ldots m_0$. The subset of the peaks with  $m_0=N$ correspond to pure spin-helix states \cite{JPA}.
}
\label{fig:phase}
\end{figure}

The MPA expression for $\rho^{(0)}$ allows for an efficient computation of local observables, such as magnetization profiles and spin current, for previously inaccessible system sizes; see Fig.~\ref{fig:profiles_scaling}. In Fig.~\ref{fig:phase} we plot the phase diagram of the spin current exhibiting high sensitivity with resonance spiking as a function
of anisotropy parameter. For a detailed analysis of the problem we refer the reader to Ref.~\cite{inPrep19}.

\subsection{Zeno limit with asymmetric dissipation rates}

Apart from the symmetric dissipative action on both boundaries, we can as well consider Lindblad problem with infinitely large, but different dissipation rates at the left and the right edge. To this end, we renormalize the Lindblad jump operators as  
\begin{align}
k_{\rm l} \rightarrow  k_{\rm l}\sqrt{\kappa},\qquad
k_{\rm r} \rightarrow k_{\rm r}/ \sqrt{\kappa},
\end{align} 
where $0<\kappa$ is the measure of the left-right asymmetry. The ratio of the effective dissipative rates is then fixed to $\kappa$.
We can now study the Zeno limit of the Lindblad master equation as a function of  $\kappa$, for the general XYZ model. 

Heuristically we observe the following remarkable fact:
nonequilibrium steady state for the XYZ model \textit{does not depend} on the asymmetry $\kappa$ of the dissipation rates, as long as
both dissipation rates go to infinitity, i.e. in the Zeno limit $\Gamma\to\infty$ (for any finite dissipation $\Gamma$ the steady state of course depends on $\kappa$). 
This property is rather exceptional and
is related to a subtle property of the Zeno effective dynamics of the XYZ model, described below.

It has been shown in \cite{2018PopkovZenoDynamics} that the effective Zeno dynamics of a quantum system is governed by (a) the dissipation 
projected Hamiltonian $H_{\cal D}$ (\ref{eq:dis_proj_ham}), (b) by a classical Markov process with rates $w_{\alpha,\beta}$, calculated using the eigenstates $\ket{\alpha}$ of $H_{\cal D}$ and some auxiliary operators $g_\mu$, calculated from the dissipator, 
\begin{align}
w_{\alpha,\beta}=\sum_\mu |\bra{\beta}g_\mu\ket{\alpha}|^2,\quad\alpha\neq\beta.
\label{eq:MarkovProcessRates}
\end{align}
For instance, under generic assumption of a non-degenerate spectrum, the bulk of the NESS density matrix is diagonal in the basis $\ket{\alpha}$, $R = \sum_\alpha p_\alpha \ket{\alpha} \bra{\alpha}$, 
where the `probability vector' $\{ p_\alpha \}$ is given as an invariant state (fixed point) of a classical Markov process
\be
\sum_{\beta \neq \alpha} w_{\beta,\alpha}\,p_\beta-p_\alpha \sum_{\beta \neq \alpha} w_{\alpha,\beta}=0.
\ee
If two systems have the same dissipation-projected Hamiltonians, and the solutions of Eq. (42) are also the same, then the Zeno limit of NESS is the same as well. This is exactly the situation we have: firstly, the dissipation 
projected Hamiltonian is just determined by the kernel of the dissipator and therefore does not depend on the asymmetry $\kappa$. It is given in Eq.~\eqref{eq:dis_proj_ham}.
Secondly, for our choice of the dissipators, the sum over $\mu$ in  \eqref{eq:MarkovProcessRates} consists of two terms, each one associated with a separate boundary; the
respective $g_{\rm l}(\kappa),g_{\rm r}(\kappa)$ can be calculated using the method developed in~\cite{2018PopkovZenoDynamics} and read
\begin{align}
\begin{aligned}
&g_{\rm l}(\kappa)= \kappa^{\frac{1}{2}}({\rm J}\vec{n}_{\rm l}'+i {\rm J}\vec{n}_{\rm l}'')\cdot \vec{\si}_1,\\
&g_{\rm r}(\kappa)= \kappa^{-\frac{1}{2}}({\rm J}\vec{n}_{\rm r}'+i {\rm J}\vec{n}_{\rm r}'')\cdot \vec{\si}_N,
\label{eq:MarkovProcessRates-1}
\end{aligned}
\end{align}
where the unit vectors $\vec{n}_\mu',\vec{n}_\mu''$, for $\mu\in\{{\rm l},{\rm r}\}$, have been introduced after Eq.~\eqref{eq:unitvectors}.

For a general ``MPA-integrable'' case we now numerically observe
\begin{align}
\left| \frac{ \bra{\beta} g_{\rm l}(1) \ket{\alpha} } { \bra{\alpha} g_{\rm l}(1) \ket{\beta} } \right|^2 &=
\left| \frac{ \bra{\beta} g_{\rm r}(1) \ket{\alpha} } { \bra{\alpha} g_{\rm r}(1) \ket{\beta} } \right|^2, \quad \forall \alpha,\beta,\label{eq:MarkovProcessRates-Proportionality}\\
w_{\alpha, \beta} w_{ \beta, \gamma} w_{\gamma, \alpha }&=
  w_{\alpha ,\gamma} w_{ \gamma, \beta} w_{\beta, \alpha }, 
\quad \forall \alpha,\beta,\gamma.\label{eq:MarkovProcessRates-Additional}
\end{align}
The properties~\eqref{eq:MarkovProcessRates-Proportionality} and~\eqref{eq:MarkovProcessRates-Additional} are very special; they hold if the bulk is described by a homogeneous XYZ (integrable) Hamiltonian. We checked, for example, that if we take a nonintegrable Hamiltonian (i.e., switch on integrability breaking terms), they are no longer satisfied.

Due to the Kolmogorov criterion~\eqref{eq:MarkovProcessRates-Additional}, the steady state probabilities $p_\alpha$ satisfy
the detailed balance condition $p_\alpha/p_\beta=w_{\beta,\alpha}/w_{\alpha,\beta}$. Then, due to Eq.~\eqref{eq:MarkovProcessRates-Proportionality}, we have
\begin{align}
\frac{p_\alpha(\kappa)}{p_\beta(\kappa)}=
 \frac{w_{\beta,\alpha}(\kappa) } {w_{\alpha,\beta}(\kappa) }=
\frac{p_\alpha(1)}{p_\beta(1)}. 
\label{MarkovProcessRates-2}
\end{align} 
Since the probabilities $p_\alpha(\kappa)$ are normalized, $\sum_\alpha p_\alpha(\kappa)=1$,
the steady state of the associated Markov process for
the asymmetric boundary driving is the same as in the case of symmetric driving: $p_\alpha(\kappa)=p_\alpha(1)$. 
Consequently, the Zeno limit of NESS will remain the same, i.e., it will be
independent of the asymmetry $\kappa$ in the dissipation at the left
and the right boundary.

\section{Discussion} 

The Sutherland equation -- divergence condition (\ref{eq:sutherland}) and the boundary equations (\ref{eq:boundary}) are two crucial ingredients in the construction of conservation laws and nonequilibrium steady states of boundary driven spin chains. Here we have
proposed a generalized, inhomogeneous Sutherland equation, in which the Lax matrices of the MPA explicitly depend on the lattice site. We have demonstrated the
applicability of the resulting MPA by generating the nonequilibrium steady state of a boundary driven XYZ spin-$1/2$ chain with strong dissipative spin-polarizing  boundary baths. Generically, our ansatz (\ref{eq:ansatz_r})  can be also used as a tool to construct nontrivial conservation laws for the open  spin chain with arbitrary nondiagonal boundary fields (\ref{eq:CommutativityRHboundary}).

The structure of constituent matrices of our ansatz (\ref{eq:ansatz_r}) is very different from that of
previously treated Lax operators, which satisfy the celebrated Yang-Baxter equation.
Besides having a site-dependent auxiliary structure, our Lax operators cannot be put into a tridiagonal form, even after all of the nonisomorphic local auxiliary spaces ${\cal A}_n$ are embedded into a joint infinite-dimensional auxiliary vector space. For example, it can be checked that our explicit representation (\ref{eq:seedXXZ1},\ref{eq:seedXXZ2}) cannot be reduced to the highest weight representation of the ${\cal U}_q(sl_2)$ quantum group symmetry of the XXZ model, which has been used to solve Lindblad equation for the longitudinal \cite{TP2011} or transverse \cite{KPS2013} dissipative boundaries. In other words, the Lax structure proposed here, seems to correspond to a new representation of the underlying symmetry algebra, in which the auxiliary space is not fixed to some ${\cal U}_q(sl_2)$ module, but rather corresponds to a ladder of linear vector spaces, transitions between which are represented by matrices of our ansatz.
Similarly, we expect that for the  anisotropic XYZ model our inhomogeneous Lax operators and nonequilibrium dissipative solutions go beyond the off-diagonal Bethe ansatz which diagonalizes the closed Hamiltonian \cite{ODBA1,ODBA2}. It is left as an open future problem to find explicit analytic expression for the inhomogeneous Lax operators in the general XYZ case,  presumably in terms of Jacobi elliptic functions.

\vspace{2mm}
We acknowledge discussions with M. Petkov\v sek and V. Romanovsky.
The work has been supported by European Research Council (ERC) through the advanced grant 694544 -- OMNES and the grant P1-0402 of Slovenian Research Agency (ARRS).
V.P. also acknowledges support by the DFG grant KL 645/20-1.

\onecolumngrid
\appendix

\section{Cancellation mechanism and the boundary equations}

In this appendix we elaborate on the boundary equations, that need to be satisfied in order for the commutation relation
\be
[H+\vec{h}_{\rm l}\cdot\vec{\sigma}_1+\vec{h}_{\rm r}\cdot\vec{\sigma}_N,R]=0
\ee
to hold for an operator $R=\Omega\Omega^\dagger$, with $\Omega=\bra{0}\L_1 \L_2\ldots \L_N\ket{\psi}$. The Hamiltonian is given by Eq.~\eqref{eq:Hamiltonian}, while the inhomogeneus Lax operators $\L_n=\vec{\sigma}_n\cdot\vec{L}_n$ satisfy the so-called divergence condition, given in Eq.~\eqref{eq:sutherland}. A straightforward application of the latter yields
\be
[H,R]=i\left(\bra{0}I_1\L_2\ldots \L_N\ket{\psi}-\bra{0}\L_1\ldots \L_{N-1}I_N\ket{\psi}\right)\Omega^\dagger+i\Omega\left(\bra{\bar{0}}I_1\L^*_2\ldots \L^*_N\ket{\bar{\psi}}-\bra{\bar{0}}\L^*_1\ldots \L^*_{N-1}I_N\ket{\bar{\psi}}\right),
\ee
where $(\bullet)^*$ denotes complex conjugation over the auxiliary space and hermitian conjugation over the physical space and $\ket{\bar{\psi}}:=(\ket{\psi})^*$. For example, for $\alpha\in{\cal J}$ we have $(\sigma_n^\alpha L^\alpha_n)^*:=(\sigma_n^\alpha)^\dagger (L_n^\alpha)^*=\sigma_n^\alpha (L_n^\alpha)^*$. Using $\mathbb{L}=\L\otimes\L^*=\sum_{\alpha,\beta\in{\cal J}}\sigma_n^\alpha \sigma_n^\beta\, L_n^\alpha \otimes (L_n^\beta)^*$, where $\otimes$ denotes the tensor product over auxiliary spaces and ordinary matrix multiplication over the physical space ${\cal H}$, we can rewrite this as
\be
[H,R]=i\bra{0, \bar{0}}(I_1\otimes \L_1^* +\L_1\otimes I_1){\mathbb L}_2\ldots {\mathbb L}_N\ket{\psi,\bar{\psi}}-i\bra{0,\bar{0}}{\mathbb L}_1\ldots{\mathbb L}_{N-1}(I_N\otimes\L^*_N+\L_N\otimes I_N)\ket{\psi,\bar{\psi}}.
\ee

On the other hand we have
\be
[\vec{h}_{\rm l}\cdot\vec{\sigma}_1,R]=\bra{0,\bar{0}}[\vec{h}_{\rm l}\cdot\vec{\sigma}_1,\mathbb{L}_1]\mathbb{L}_2\ldots\mathbb{L}_N\ket{\psi,\bar{\psi}},\quad
[\vec{h}_{\rm r}\cdot\vec{\sigma}_N,R]=\bra{0,\bar{0}}\mathbb{L}_1\ldots\mathbb{L}_{N-1}[\vec{h}_{\rm r}\cdot\vec{\sigma}_N,\mathbb{L}_N]\ket{\psi,\bar{\psi}},
\ee
where the commutators can be explicitly rewritten as
\bea
&&[\vec{h}_{\rm l}\cdot\vec{\sigma}_1,\mathbb{L}_1]=\sum_{\alpha,\beta,\gamma\in{\cal J}}h_{\rm l}^\alpha\,L^\beta_1\otimes(L^\gamma_1)^*\,[\sigma_1^\alpha,\sigma_1^\beta \sigma_1^\gamma]=\sum_{\alpha,\beta,\gamma,\delta\in{\cal J}}h_{\rm l}^\alpha\,L^\beta_1\otimes(L^\gamma_1)^*\,i\,\varepsilon_{\beta,\gamma,\delta}\,[\sigma_1^\alpha,\sigma_1^\delta]=\notag\\[1em]
&&=\sum_{\alpha,\beta,\gamma,\delta,\omega\in{\cal J}}2\,h_{\rm l}^\alpha\,L^\beta_1\otimes(L^\gamma_1)^*\,\varepsilon_{\delta,\beta,\gamma}\,\varepsilon_{\delta,\alpha,\omega}\,\sigma_1^\omega=\sum_{\alpha,\beta,\gamma,\omega\in{\cal J}}2\,h_{\rm l}^\alpha\,L^\beta_1\otimes(L^\gamma_1)^*\,(\delta_{\beta,\alpha}\delta_{\gamma,\omega}-\delta_{\beta,\omega}\delta_{\gamma,\alpha})\,\sigma_1^\omega=\notag\\[1em]
&&=\sum_{\alpha,\gamma\in{\cal J}}2\,h_{\rm l}^{\alpha}\,L_1^\alpha\otimes(L_1^\gamma)^*\,\sigma_1^\gamma-\sum_{\alpha,\beta\in{\cal J}}2\,h_{\rm l}^{\alpha}\,L_1^\beta\otimes(L_1^\alpha)^*\,\sigma_1^\beta=2\,(\vec{h}_{\rm l}\cdot\vec{L}_1)\otimes\L^*_1-2\,\L_1\otimes(\vec{h}_{\rm l}\cdot \vec{L}^*_1)
\eea
and similarly $[\vec{h}_{\rm r}\cdot\vec{\sigma}_N,\mathbb{L}_N]=2\,(\vec{h}_{\rm r}\cdot\vec{L}_N)\otimes\L^*_N-2\,\L_N\otimes(\vec{h}_{\rm r}\cdot \vec{L}^*_N)$. Putting everything together, we get
\be
\begin{gathered}
\big[H+\vec{h}_{\rm l}\cdot\vec{\sigma}_1+\vec{h}_{\rm r}\cdot\vec{\sigma}_N,R\big]=\bra{0,\bar{0}}\left([2\,\vec{h}_{\rm l}\cdot\vec{L}_1+iI_1]\otimes \L_1^* -\L_1\otimes [2\,\vec{h}_{\rm l}\cdot \vec{L}^*_1-iI_1]\right)\mathbb{L}_2\ldots\mathbb{L}_N\ket{\psi,\bar{\psi}}+\\
+\bra{0,\bar{0}}\mathbb{L}_1\ldots\mathbb{L}_{N-1}\left([2\,\vec{h}_{\rm r}\cdot\vec{L}_N-iI_N]\otimes\L^*_N-\L_N\otimes[2\,\vec{h}_{\rm r}\cdot \vec{L}^*_N+iI_N]\right)\ket{\psi,\bar{\psi}}.
\end{gathered}
\ee
If the boundary equations $\bra{0}[2\,\vec{h}_{\rm l}\cdot\vec{L}_1+iI_1]=0$ and $[2\,\vec{h}_{\rm r}\cdot\vec{L}_N-iI_N]\ket{\psi}=0$ are satisfied, the operator $R$ commutes with the Hamiltonian $H+\vec{h}_{\rm l}\cdot\vec{\sigma}_1+\vec{h}_{\rm r}\cdot\vec{\sigma}_N$.

\section{Proof of the ansatz in the XXZ case}

In the XXZ case, the tensor of anisotropic spin-spin interactions becomes ${\rm J}={\rm diag}(1,1,\cos\gamma)$. Writing $L^x_n =\frac{1}{2}(L^+_n + L^-_n)$ and $L^y_n = \frac{1}{2i}(L^-_n - L^+_n)$,
the discrete spatial Landau-Lifshitz equations given by Eq.~\eqref{eq:recurrence} hold, if
\begin{align}
\begin{aligned}
&L_n^+L_{n+1}^--L_n^- L_{n+1}^+=i\,L_n^z I^{}_{n+1}, & \quad &L_n^+L_{n+1}^--L_n^- L_{n+1}^+=i\,I^{}_n L_{n+1}^z,\\[1em]
&L_n^z L_{n+1}^+-\cos\gamma\,L_n^+ L_{n+1}^z=\frac{i}{2}\,L_n^+I^{}_{n+1}, & \quad &\cos\gamma\,L_n^z L_{n+1}^+- L_n^+ L_{n+1}^z=\frac{i}{2}\,I^{}_{n} L_{n+1}^+\\[1em]
&L_n^z L_{n+1}^--\cos\gamma\,L_n^- L_{n+1}^z=-\frac{i}{2}\,L_n^-I^{}_{n+1}, & \quad & \cos\gamma\,L_n^z L_{n+1}^-- L_n^- L_{n+1}^z=-\frac{i}{2}\,I^{}_{n} L_{n+1}^-.
\label{sup:relations}
\end{aligned}
\end{align}
Our goal in this appendix is, to show that the ansatz
\be
\begin{gathered}
L^z_n = \sum_{k=0}^{n-1} \ket{k}\!\bra{k+1},\qquad L^\pm_n= \pm \eta^{\mp 1}\sum_{k=0}^{n-1}\sum_{l=0}^{n}\left(\frac{\pm i}{2 \cos\gamma}\right)^{k-l+1}\!\!\!M_{n;k,l}\ket{k}\!\bra{l},\\
M_{n;k,l} = \frac{(\xi-\xi^{-1})\, P_{n,k+1,l}(\cos\gamma)}{(\xi+\xi^{-1})\,\sin\gamma}-\frac{2P_{n,k+1,l-1}(\cos\gamma)}{(\xi+\xi^{-1})\,\cos\gamma},\qquad P_{n,k,l}(x)=\sum_{s=0}^l (-1)^s {n-k\choose s}{n-s-1\choose l-s}\,x^{n-2s}
\label{sup:polynomials}
\end{gathered}
\ee
satisfies algebraic relations~\eqref{sup:relations}. This will be done in two parts. Firstly, we will discuss three lemmas which will facilitate the proof of the relations themselves.
The latter will be presented in the second part.

\subsection{Lemmas}

\begin{lem}
Polynomials given in~\eqref{sup:polynomials}, satisfy the following recurrence relations
\begin{align}
\begin{aligned}
P_{n,k,l}(x)=x\,[P_{n-1,k-1,l-1}(x)+P_{n-1,k-1,l}(x)],\\
P_{n,k,l}(x)=x\,[P_{n+1,k+1,l+1}(x)-P_{n+1,k,l+1}(x)].\notag
\end{aligned}
\end{align}
\end{lem}
\proof This is a simple consequence of the Pascal rule for the binomial coefficients. \endproof

\begin{rem}
Note, that the recurrence relations hold irrespective of what integer $l$ is. For example, we have $P_{n,k,0}(x)=x^n$ and $P_{n,k,l}=0$ for $l<0$, which is consistent with the relations.
\end{rem}

\begin{lem}
Binomial coefficients satisfy relations
\begin{align}
\sum_{s=0}^{n-1}(-1)^s{n-t-1\choose s-t}{n-s\choose t'}=(-1)^t\,\big(\delta_{t',n-t}+\delta_{t',n-t-1}\big),\notag
\end{align}
for $0\le t \le n-1$, $t'\in \ZZ$ and
\begin{align}
\sum_{s=0}^{n}(-1)^s{n-t-1\choose s-t-1}{n-s\choose t'}=(-1)^{t+1}\delta_{t',n-t-1},\notag
\end{align}
for $-1\le t\le n-1$, $t'\in\ZZ$.
\end{lem}
\proof The first relation is
\begin{align}
\begin{gathered}
\sum_{s=0}^{n-1}(-1)^s{n-t-1\choose s-t}{n-s\choose t'}=\sum_{s=0}^{n-1}(-1)^s\Big\{{n-t-1\choose n-s-1}{n-s-1\choose t'-1}+{n-t-1\choose n-s-1}{n-s-1\choose t'}\Big\}=\\
=\sum_{s'=0}^{n-1}(-1)^{n-s'-1}{n-t-1\choose s'}{s'\choose t'-1}+\sum_{s'=0}^{n-1}(-1)^{n-s'-1}{n-t-1\choose s'}{s'\choose t'}=\\
=\sum_{s'=t'-1}^{n-t-1}(-1)^{n-s'-1}{n-t-1\choose s'}{s'\choose t'-1}+\sum_{s'=t'}^{n-t-1}(-1)^{n-s'-1}{n-t-1\choose s'}{s'\choose t'}=\\
=(-1)^t\,\big(\delta_{t',n-t}+\delta_{t',n-t-1}\big).\notag
\end{gathered}
\end{align}
In the last equality we have used one of the standard binomial sum identities: $\sum_{s=m}^n(-1)^{n-s}{n\choose s}{s\choose m}=\delta_{n,m}$. To do so, we have truncated the sums, $\sum_{s'=0}^{n-1}\to \sum_{s'=t'-1}^{n-t-1}$ and $\sum_{s'=0}^{n-1}\to \sum_{s'=t'}^{n-t-1}$, respectively. This is possible even for $t'\le0$. In this case, the first sum will vanish, since ${a\choose b}=0$ if $b<0$. On the other hand, we can only change the upper bound from $n-1$ to $n-t-1$ if $t\ge0$.

The second relation is
\begin{align}
\begin{gathered}
\sum_{s=0}^{n}(-1)^s{n-t-1\choose s-t-1}{n-s\choose t'}=\sum_{s=0}^{n}(-1)^s{n-t-1\choose n-s}{n-s\choose t'}=\\
=\sum_{s'=0}^{n}(-1)^{n-s'}{n-t-1\choose s'}{s'\choose t'}=\sum_{s'=t'}^{n-t-1}(-1)^{n-s'}{n-t-1\choose s'}{s'\choose t'}=(-1)^{t+1}\delta_{t',n-t-1}.\notag
\end{gathered}
\end{align}
Again, we have used the identity $\sum_{s=m}^n(-1)^{n-s}{n\choose s}{s\choose m}=\delta_{n,m}$, after truncating the sum according to $\sum_{s'=0}^n\to\sum_{s'=t'}^{n-t-1}$. This is possible even for $t=-1$ and $t'\le 0$. \endproof

\begin{lem}
For $0\le k\le n-1$ polynomials given by \eqref{sup:polynomials}, satisfy the relations
\begin{align}
&\sum_{s=0}^{n}(-1)^s P_{n,k+1,s}(x)P_{n+1,s+1,l}(x)=(-1)^k\,\big(\delta_{k,l}x^3+\delta_{k,l-1} (x^3-x)\big),\notag\\
&\sum_{s=0}^{n}(-1)^s P_{n,k+1,s-1}(x)P_{n+1,s+1,l}(x)=(-1)^{k+1}\,x^3\,\big(\delta_{k,l}+\delta_{k,l-1}\big).\notag
\end{align}
\end{lem}

\proof We start by proving the first relation. We write out the left hand side:
\begin{align}
\begin{gathered}
\sum_{s=0}^{n}(-1)^s P_{n,k+1,s}(x)P_{n+1,s+1,l}(x)=\\
=\sum_{s=0}^{n}(-1)^s\sum_{t=0}^{s}(-1)^{t}{n-k-1\choose t}{n-t-1\choose s-t}x^{n-2t}\sum_{t'=0}^{l}(-1)^{t'}{n-s\choose t'}{n-t'\choose l-t'}x^{n-2t'+1}.\notag
\end{gathered}
\end{align}
Since ${a\choose b}=0$ for $a<b$ or $b<0$, we can truncate the sum over $s$ at $n-1$ and extend sums over $t$ and $t'$ up to $n-1$ and $n$, respectively. We get
\begin{align}
\begin{gathered}
=\sum_{t'=0}^{n}\sum_{t=0}^{n-1}(-1)^{t+t'}x^{2n-2t-2t'+1}{n-k-1\choose t}{n-t'\choose n-l}\underbrace{\sum_{s=0}^{n-1}(-1)^s{n-t-1\choose s-t}{n-s\choose t'}}_{(-1)^t\,(\delta_{t',n-t}+\delta_{t',n-t-1})},\notag
\end{gathered}
\end{align}
which, after using Lemma 2, becomes
\begin{align}
\begin{gathered}
=\sum_{t=0}^{n-1}(-1)^{n-t}{n-k-1\choose t}{t\choose n-l}x+\sum_{t=0}^{n-1}(-1)^{n-t-1}{n-k-1\choose t}{t+1\choose n-l}x^3=\\
=\sum_{t=n-l}^{n-k-1}(-1)^{n-t}{n-k-1\choose t}{t\choose n-l}x+\sum_{t=0}^{n-1}(-1)^{n-t-1}{n-k-1\choose t}\Big\{{t\choose n-l}+{t\choose n-l-1}\Big\}x^3=\\
=(-1)^{k+1}\delta_{k,l-1}x+\sum_{t=n-l}^{n-k-1}(-1)^{n-t-1}{n-k-1\choose t}{t\choose n-l}x^3+\sum_{t=n-l-1}^{n-k-1}(-1)^{n-t-1}{n-k-1\choose t}{t\choose n-l-1}x^3=\\
=(-1)^{k+1}\delta_{k,l-1}x+(-1)^k\delta_{k,l-1}x^3+(-1)^k\delta_{k,l}x^3=(-1)^k\Big(\delta_{k,l}x^3+\delta_{k,l-1}(x^3-x)\Big).\notag
\end{gathered}
\end{align}
To produce the Kronecker deltas via identity $\sum_{s=m}^n(-1)^{n-s}{n\choose s}{s\choose m}=\delta_{n,m}$, we had to truncate the sum over $t$ at $n-k-1$. This is allowed by the assumption $k\ge 0$.

The second relation is even simpler to prove, again starting by writing out the left hand side:
\begin{align}
\begin{gathered}
\sum_{s=0}^{n}(-1)^s P_{n,k+1,s-1}(x)P_{n+1,s+1,l}(x)=\\
=\sum_{s=0}^{n}(-1)^s\sum_{t=0}^{s-1}(-1)^t{n-k-1\choose t}{n-t-1\choose s-t-1}x^{n-2t}\sum_{t'=0}^{l}(-1)^{t'}{n-s\choose t'}{n-t'\choose l-t'}x^{n-2t'+1}.\notag
\end{gathered}
\end{align}
Since ${a\choose b}=0$ for $b>a$ or $b<0$, we can extend the sum over $t$ up to the maximum $s-1=n-1$. The sum over $t'$ can be extended up to $n$. We then get
\begin{align}
\begin{gathered}
=\sum_{t'=0}^{n}\sum_{t=0}^{n-1}(-1)^{t+t'}x^{2n-2t-2t'+1}{n-k-1\choose t}{n-t'\choose n-l}\underbrace{\sum_{s=0}^{n}(-1)^s{n-t-1\choose s-t-1}{n-s\choose t'}}_{(-1)^{t+1}\delta_{t',n-t-1}}\notag
\end{gathered}
\end{align}
and using Lemma 2,
\begin{align}
\begin{gathered}
=\sum_{t=0}^{n-1}(-1)^{n-t}{n-k-1\choose t}{t+1\choose n-l}x^3=\sum_{t=0}^{n-1}(-1)^{n-t}{n-k-1\choose t}\Big\{{t\choose n-l}+{t\choose n-l-1}\Big\}x^3=\\
=\sum_{t=n-l}^{n-k-1}(-1)^{n-t}{n-k-1\choose t}{t\choose n-l}x^3+\sum_{t=n-l-1}^{n-k-1}(-1)^{n-t}{n-k-1\choose t}{t\choose n-l-1}x^3=\\
=(-1)^{k+1}\,x^3\,(\delta_{k,l-1}+\delta_{k,l}).\notag
\end{gathered}
\end{align}
This completes the proof of the three lemmas. \endproof

\begin{rem}
The polynomial relations from Lemma 3 are trivially satisfied even for $l< 0$, since then $P_{n,k,l}=0$.
\end{rem}

\subsection{Proof of the algebraic relations}

We will now finally prove, that the ansatz \eqref{sup:polynomials} satisfies the algebraic relations \eqref{sup:relations}, using the Lemmas 1 and 3 from the previous subsection.

\begin{cor}
Relations $L_n^+L_{n+1}^--L_n^- L_{n+1}^+=i\,L_n^z I^{}_{n+1}$ and $L_n^+L_{n+1}^--L_n^- L_{n+1}^+=i\, I^{}_{n}L_{n+1}^z$ are satisfied.
\end{cor}
\proof Since they are equivalent, we will only prove the first one. Explicitly, the first relation reads
\begin{align}
\sum_{k=0}^{n-1}\sum_{l=0}^{n+1}\left(\frac{(-1)^k-(-1)^l}{4\cos^2\gamma}\left(\frac{i}{2\cos\gamma}\right)^{k-l}A_{k,l}\right)\ket{k}\bra{l}=\sum_{k=0}^{n-1}\sum_{l=0}^{n+1}\left(i\,\delta_{k,l-1}\right)\ket{k}\bra{l},\label{sup:component_form}
\end{align}
where
\begin{align}
\begin{gathered}
A_{k,l}=\sum_{s=0}^{n}(-1)^s M_{n;k,s}M_{n+1;s,l}=\frac{1}{\sin^2\gamma}\sum_{s=0}^{n}(-1)^s P_{n,k+1,s}(\cos\gamma)P_{n+1,s+1,l}(\cos\gamma)+\\
+\frac{4}{(\xi+\xi^{-1})^2}\sum_{s=0}^{n}(-1)^s\Big(\frac{1}{\cos^2\gamma}P_{n,k+1,s-1}(\cos\gamma)P_{n+1,s+1,l-1}(\cos\gamma)-\frac{1}{\sin^2\gamma}P_{n,k+1,s}(\cos\gamma)P_{n+1,s+1,l}(\cos\gamma)-\\
-\frac{\xi-\xi^{-1}}{2\cos\gamma\sin\gamma}\left[P_{n,k+1,s}(\cos\gamma)P_{n+1,s+1,l-1}(\cos\gamma)+P_{n,k+1,s-1}(\cos\gamma)P_{n+1,s+1,l}(\cos\gamma)\right]\Big).\notag
\end{gathered}
\end{align}
Because of the prefactor $(-1)^k-(-1)^l$, only the cases where $k-l$ is an odd integer need to be checked. Since there is no $\xi$-dependence on the right hand side of the relation~\eqref{sup:component_form}, the second sum in $A_{k,l}$ should be zero. Note, that we can use Lemma 3 in all of the terms of the matrix element $A_{k,l}$. This gives
\begin{align}
\begin{gathered}
A_{k,l}=\frac{1}{\sin^2\gamma}(-1)^k\,\big(\delta_{k,l-1} (\cos^3\gamma-\cos\gamma)+\delta_{k,l}\cos^3\gamma\big)+\\
+\frac{4}{(\xi+\xi^{-1})^2}\Big(
(-1)^{k+1}\cos\gamma\big(\delta_{k,l-1}+\delta_{k,l-2}\big)+(-1)^{k}\cos\gamma\,\delta_{k,l-1}-(-1)^{k}\delta_{k,l}\frac{\cos^3\gamma}{\sin^2\gamma}-\\
-\frac{\xi-\xi^{-1}}{2\cos\gamma\sin\gamma}\big[(-1)^k\big(\delta_{k,l-2} (\cos^3\gamma-\cos\gamma)+\delta_{k,l-1}\cos^3\gamma\big)+(-1)^{k+1}\cos^3\gamma\,(\delta_{k,l}+\delta_{k,l-1})\big]\Big).\notag
\end{gathered}
\end{align}
For odd $k-l$ it becomes $A_{k,l}=(-1)^{k+1}\cos\gamma\,\delta_{k,l-1}$. Using this result we see that \eqref{sup:component_form} is indeed satisfied. \endproof

\begin{cor}
Relations $L_n^z L_{n+1}^\pm-\cos\gamma\,L_n^\pm L_{n+1}^z=\pm\frac{i}{2}\,L_n^\pm I^{}_{n+1}$ are satisfied.
\end{cor}

\proof Explicitly, they are both equivalent to
\begin{align}
\sum_{k=0}^{n-1}\sum_{l=0}^{n+1}\left(\frac{\pm i}{2\cos\gamma}\right)^{k-l+2}
\left(M_{n+1;k+1,l}-\cos\gamma\, M_{n;k,l-1}\right)\ket{k}\bra{l}=\sum_{k=0}^{n-1}\sum_{l=0}^{n+1}\left(\frac{\pm i}{2\cos\gamma}\right)^{k-l+2}(\cos\gamma\,M_{n;k,l})\ket{k}\bra{l},\notag
\end{align}
where we note $M_{n;k,l-1}=0$, for $l=0$. They are obviously satisfied by courtesy of the first polynomial recurrence in Lemma~1. \endproof

\begin{cor}
Relations $\cos\gamma\,L_n^z L_{n+1}^\pm-L_n^\pm L_{n+1}^z=\pm\frac{i}{2}\, I^{}_{n} L_{n+1}^\pm$ are satisfied.
\end{cor}

\proof Explicitly, they read
\begin{align}
\sum_{k=0}^{n-1}\sum_{l=0}^{n+1}\left(\frac{\pm i}{2\cos\gamma}\right)^{k-l+2}
\left(\cos\gamma\, M_{n+1;k+1,l}-M_{n;k,l-1}\right)\ket{k}\bra{l}=\sum_{k=0}^{n-1}\sum_{l=0}^{n+1}\left(\frac{\pm i}{2\cos\gamma}\right)^{k-l+2}(\cos\gamma\,M_{n+1;k,l})\ket{k}\bra{l}.\notag
\end{align}
Again, note $M_{n;k,l-1}=0$, for $l=0$.  These relations are satisfied due to the second polynomial recurrence in Lemma~1. \endproof

\twocolumngrid

\end{document}